\documentclass{aa}  

\usepackage{graphicx}
\usepackage{txfonts}
\usepackage[bookmarks=false,         
     pdfnewwindow=true,      
     colorlinks=true,    
     linkcolor=blue,     
     citecolor=blue,     
     filecolor=blue,  
     urlcolor=blue,      
     final=true
     ]{hyperref}
\begin{document}


\title{Sapaki: Galactic O3If* star possibly born in isolation}

   \author{M. S. Zarricueta Plaza
          \inst{1}
          \and
          A. Roman-Lopes
          \inst{1}
          \and
          D. Sanmartim
          \inst{2,3}
          }

   \institute{Departamento de Astronomía, Universidad de La Serena, Avenida Juan                    Cisternas 1200, La Serena, Chile \\
        \email{matiassebastian.zarricueta@userena.cl}
         \and
             Carnegie Observatories, Las Campanas Observatory, Casilla 601, La Serena, Chile
         \and
            Rubin Observatory Project Office, 950 N. Cherry Ave., Tucson, AZ 85719, USA
             }

   \date{Received January 06, 2023; accepted May 04, 2023}




 
  \abstract
   {The study of high-mass stars found to be isolated in the field of the Milky Way may help to probe the feasibility of the core-accretion mechanism in the case of massive star formation. The existence of truly isolated stars may efficiently probe the possibility that individual massive stars can be born in isolation.}
   {We observed WR67a (hereafter Sapaki), an O3If* star that appears to be isolated close to the center of a well-developed giant cavity that is aptly traced by 8.0 $\mu$m hot dust emission.}
   {We acquired medium-resolution ($R=4100$) and moderate signal-to-noise ($S/N = 95$ at 4500 \r{A}) spectra for Sapaki in the range of $3800-10500$ \r{A} with the Magellan Echellette (MagE) at Las Campanas Observatory. We computed the line-of-sight total extinctions. Additionally, we restricted its heliocentric distance by using a range of different estimators. Moreover, we measured its radial velocity from several lines in its spectrum. Finally, we analyzed its proper motions from Gaia to examine its possible runaway status.}
   {The star has been classified as having the spectral type O3If* given its resemblance to standard examples of the class. In addition, we found that Sapaki is highly obscured, reaching a line-of-sight extinction value of $A_{V} = 7.87$. We estimated the heliocentric distance to be in the range of $d = 4-7$ kpc. We also estimated its radial velocity to be $V_{r} = -34.2 \pm 15.6$ km/s. We may also discard its runaway status solely based on its 2D kinematics. Furthermore, by analyzing proper motions and parallaxes provided by Gaia, we found only one other star with compatible measurements.}
   {Given its apparent non-runaway status and the absence of clustering, Sapaki appears to be a solid candidate for isolated high-mass star formation in the Milky Way.}


   \maketitle
%

\section{Introduction}

   Massive stars are key actors in the dynamical and chemical evolution history of all types of galaxies, including our own Milky Way (Tielens \citeyear{T2005}). During their relatively short lifetimes, they can significantly alter the local interstellar medium (ISM) through intense feedback and supernova explosions (Fierlinger et al. \citeyear{Fetal2016}). Furthermore, they are prime sources of Lyman continuum ionizing radiation, which interacts with the ISM to sculpt bubbles and H {\small II} regions (Krause et al. \citeyear{Ketal2013}). It has been well established that the majority of high-mass stars are formed in the core of their host clusters, possibly forming binary or multiple stellar systems (Sana et al. \citeyear{Setal2013}). However, some studies have found these stars in apparent isolation in the field (e.g. de Wit et al. \citeyear{dWetal2005}; Gvaramadze et al. \citeyear{Getal2011}; Roman-Lopes \citeyear{RL2013}), which is in defiance of current theories of massive star formation that require physical conditions that are found only in the densest regions of giant molecular clouds (McKee \& Ostriker \citeyear{MO2007}; Zinnecker \& Yorke \citeyear{ZY2007}). In fact, most isolated high-mass objects are runaway stars that were ejected from their parental clusters either due to dynamical interaction with other massive companions or supernova explosions in close binaries (e.g. Perets \& Šubr \citeyear{PS2012}; Dorigo-Jones et al. \citeyear{DJetal2020}; Sana et al. \citeyear{Setal2022}). For example, in the Milky Way, the cases of Bajamar (O3.5III(f*)) and Toronto (O6.5V((f))z) offer examples of stars that were ejected from the Bermuda cluster at approximately the same time, causing a rapid orphanization of the cluster (Maíz-Apellániz et al. \citeyear{MAetal2022}). Also, WR20aa and WR20c are two OIf*/WN stars located in the vicinity of Westerlund 2, whose radial vector intercepts at the center of the cluster where the stellar density reach its maximum (Roman-Lopes et al. \citeyear{RLetal2011}). Apart from Milky Way examples, in the Large Magellanic Cloud, we have the case of VFTS 16, an O2IIIf* star that was ejected from R136, as well as VFTS 72, a fast-moving O2III-V(n)((f*)) star that was also probably ejected from its birthplace in the vicinity of R136 (Lennon et al. \citeyear{Letal2018}). These results support the idea that massive stars are, in fact, born in clusters, because the birthplace of isolated stars could be traced back to a nearby cluster (de Wit et al. \citeyear{dWetal2005}; Zinnecker \& Yorke \citeyear{ZY2007}; Gvaramadze et al. \citeyear{Getal2012}).

    In this sense, in order to prove that a massive star was formed in isolation, it has to be very young, with accurate proper motions and radial velocity measurements; it should also be imaged with exquisite angular resolution instruments (e.g., Gvaramadze et al. \citeyear{Getal2012}; Stephens et al. \citeyear{Setal2017}; Kalari et al. \citeyear{Ketal2019}). This last point is transcendental, since, thanks to near-infrared (NIR) diffraction-limited images, we can resolve the low-mass end of the initial mass function (IMF) around massive stars, especially when we search for them deeply immersed into the Galactic disk. Consequently, the detection of even a single very early O-type star represents an achievement in itself, since the scarcity of these objects is partly responsible for the poorly constrained stellar parameters, such as their initial and current masses (e.g., Walborn et al. \citeyear{Wetal2002}; Roman-Lopes \citeyear{RL2011}; Wu et al. \citeyear{Wetal2014}; Vink et al. \citeyear{Vetal2015}; Maíz-Apellániz et al. \citeyear{MAetal2017}). 

\begin{figure}
      \centering
      \includegraphics[width=3in]{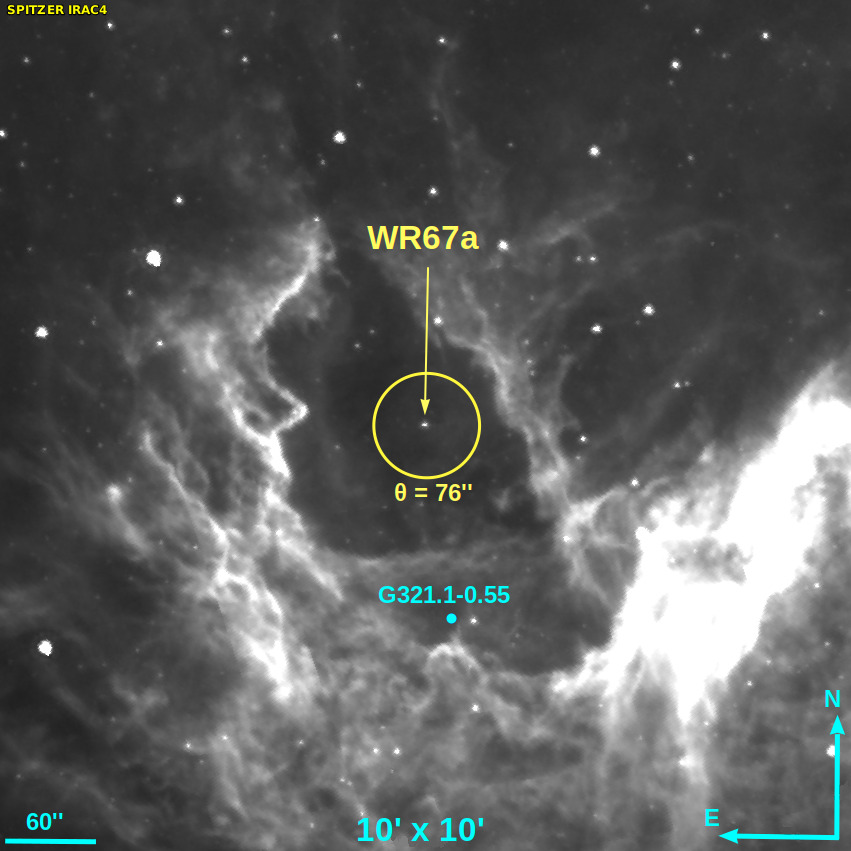}
      \caption{ $10\times10$ arcmin$^{2}$ Spitzer 8.0 $\mu$m image toward Sapaki (former WR67a). The yellow circle is centered on the star and to the same angular diameter shown in Fig. \ref{fig:Sapaki_acq} ($\theta = 76''$). The position corresponding to the coordinates of the H {\small II} region G321.1-0.55 (Caswell \& Haynes \citeyear{CH1987}) is indicated by the cyan dot.}
      \label{fig:Sapaki_image}
    \end{figure}

    In this work, we study WR67a (hereinafter referred to as Sapaki\footnote{The word translates into solitary from the Quechua language, which is spoken by several indigenous peoples in the North of Chile.}), an apparently isolated and highly-reddened luminous star located close to the Galactic midplane ($b = -0.52^{\circ}$, see Fig. \ref{fig:Sapaki_image} and Table \ref{tab:Sapaki_astrophot}). Based on the detection of selective near-infrared He, N, and C emission lines, Roman-Lopes (\citeyear{RL2011}) classified it as a WN star and proposed two scenarios: 1) Sapaki could be either a very luminous ($M_{Ks}=-6.95$ mag) and distant ($d = 10.5\pm2.1$ kpc) WN6h star or 2) a much closer binary system ($d = 3.4\pm0.8$ kpc) whose main component is a less luminous ($M_{Ks} = -4.41$) weakly lined WN6 star with emission that has diminished due to the presence of a line-of-sight stellar companion. Curiously, Sapaki appears to be nearly at the center of a well-developed cavity that is distinguished in 8.0 $\mu$m images due to hot-dust emission (see Fig. \ref{fig:Sapaki_image}). Furthermore, Caswell \& Haynes (\citeyear{CH1987}) detected, from a radio recombination line survey, an H {\small II} region (G321.1-0.55) located at $\sim2.2'$ to the south of Sapaki with a measured radial velocity of $V_{r} = -56$ km/s. The region G321.1-0.55 was also catalogued as an H {\small II} region by Anderson et al. (\citeyear{Aetal2014}) based on the Wide-field Infrared Survey Explorer (WISE) satellite. It is very likely that Sapaki is the main body responsible for the ionization of the H {\small II} region, given the colossal production of Lyman continuum photons coming from very early O-types, OIf*/WN, and Wolf-Rayet stars (Steinberg et al. \citeyear{Setal2003}; Smith \citeyear{S2006}). In Section \ref{sec:observations}, we describe the observational data and the reduction procedures. In Section \ref{sec:results}, we present the results and discussion and in Section \ref{sec:conclusion}, we summarize the main conclusions of our work.

    \begin{table}[t]
      \centering
      \begin{tabular}{lc}
      \hline
      ID & Sapaki \\ [1pt] \hline
      $\alpha_{2000}$ ($^{\text{h}}$ $^{\text{m}}$ $^{\text{s}}$) & 15:16:36.96 \\ [1pt]
      $\delta_{2000}$ ($^{\circ}$ $'$ $''$) & -58:09:58.7 \\ [1pt] 
      $l$ ($^{\circ}$) & $321.125$ \\ [1pt]
      $b$ ($^{\circ}$) & $-0.518$ \\ [1pt] \hline
      $\bar{\omega}$ (mas) & 0.19 $\pm$ 0.04 \\ [1pt]
      $\mu_{\alpha}$ (mas/yr) & -5.01 $\pm$ 0.04 \\ [1pt]
      $\mu_{\delta}$ (mas/yr) & -3.62 $\pm$ 0.05 \\ [1pt] \hline
      $B$ (mag) & 17.26 $\pm$ 0.10 \\ [1pt]
      $V$ (mag) & 14.90 $\pm$ 0.05 \\ [1pt]
      $G$ (mag) & 13.42 $\pm$ 0.01 \\ [1pt]
      $J$ (mag) & 9.86 $\pm$ 0.02 \\ [1pt]
      $H$ (mag) & 9.18 $\pm$ 0.03 \\ [1pt]
      $K_{s}$ (mag) & 8.82 $\pm$ 0.02 \\ [1pt]
      $W1$ (mag) & 8.40 $\pm$ 0.02 \\ [1pt]
      $W2$ (mag) & 8.22 $\pm$ 0.02 \\ [1pt]
      [8.0] (mag) & 7.67 $\pm$ 0.05 \\ [1pt]
      \hline
      \end{tabular}
      \caption{Literature data for Sapaki. Coordinates, parallaxes, proper motions, and G-band photometry were retrieved from Gaia Data Release 3 (Gaia Collaboration \citeyear{Gaia2022}). The $B-$ and $V$-band photometry values were taken from the TESS Input Catalog V8.2 (Paegert et al. \citeyear{Petal2021}). $J$-, $H$- and $K_{s}$-band photometry was taken from the 2MASS Point Source Catalog (Skrutskie et al. \citeyear{Setal2006}). W1-band and W2-band photometry values were taken from the WISE All-Sky Data Release (Cutri et al. \citeyear{Cetal2012}), while the [8.0] $\mu$m photometry was taken from the GLIMPSE Source Catalog (Benjamin et al. \citeyear{Betal2003})}
      \label{tab:Sapaki_astrophot}
    \end{table}

\section{Observations and data reduction} \label{sec:observations}

   The optical spectral data for Sapaki were obtained with the Magellan Echellete (MagE) mounted on the 6.5-m Magellan Baade telescope on July 31, 2022. The star was observed as part of a Chilean Telescope Allocation Committee (CNTAC) project, whereby the observing run was assigned with the code Q4FEM5BG. The observations were carried out under dark time, seeing conditions in the range $0.78-1.20''$ (mean of $0.93''$), and airmass values between $1.15-1.87$ (mean of $1.36$). The spectrograph was configured with the 1.0'' slit and an $1\times1$ binning, resulting in a mean spectral resolution of $R=4100$ and spectral coverage of $\sim3800-10500$ \r{A} across 11 spectral orders. We performed 16 exposures of 1200 seconds each as Sapaki is very faint at blue wavelengths ($B = 17.26$ mag, see Table \ref{tab:Sapaki_astrophot}). The resultant signal-to-noise level was $S/N=95$ at 4500 \r{A} if we assume a Poisson noise.

    The raw frames were reduced by using the MagE pipeline, which is part of the Carnegie Python distribution (CarPy\footnote{\href{https://code.obs.carnegiescience.edu/mage-pipeline}{https://code.obs.carnegiescience.edu/mage-pipeline}}). In addition, the data were later retouched with standard IRAF\footnote{\href{http://iraf.noao.edu/}{http://iraf.net/}} routines. At first, the MagE pipeline reads a database file containing the relevant observational parameters in order to generate the appropriate pipeline recipes. Subsequently, the calibration pipelines are run to generate the normalized flats for the red and blue sides of each spectrum separately. A set of Xe-Flash frames are then acquired to map the order edges, and a set of ThAr lamp frames are also obtained to estimate the line curvature and to perform the wavelength calibration. Afterward, the science pipelines are run in order to properly reduce the data and to perform the one-dimensional (1D) extraction for each order. Later on, the images  are\ processed using IRAF tasks from the \texttt{onedspec} package. 

\begin{figure*}[htbp!]
      \centering
      \includegraphics[width=6.5in]{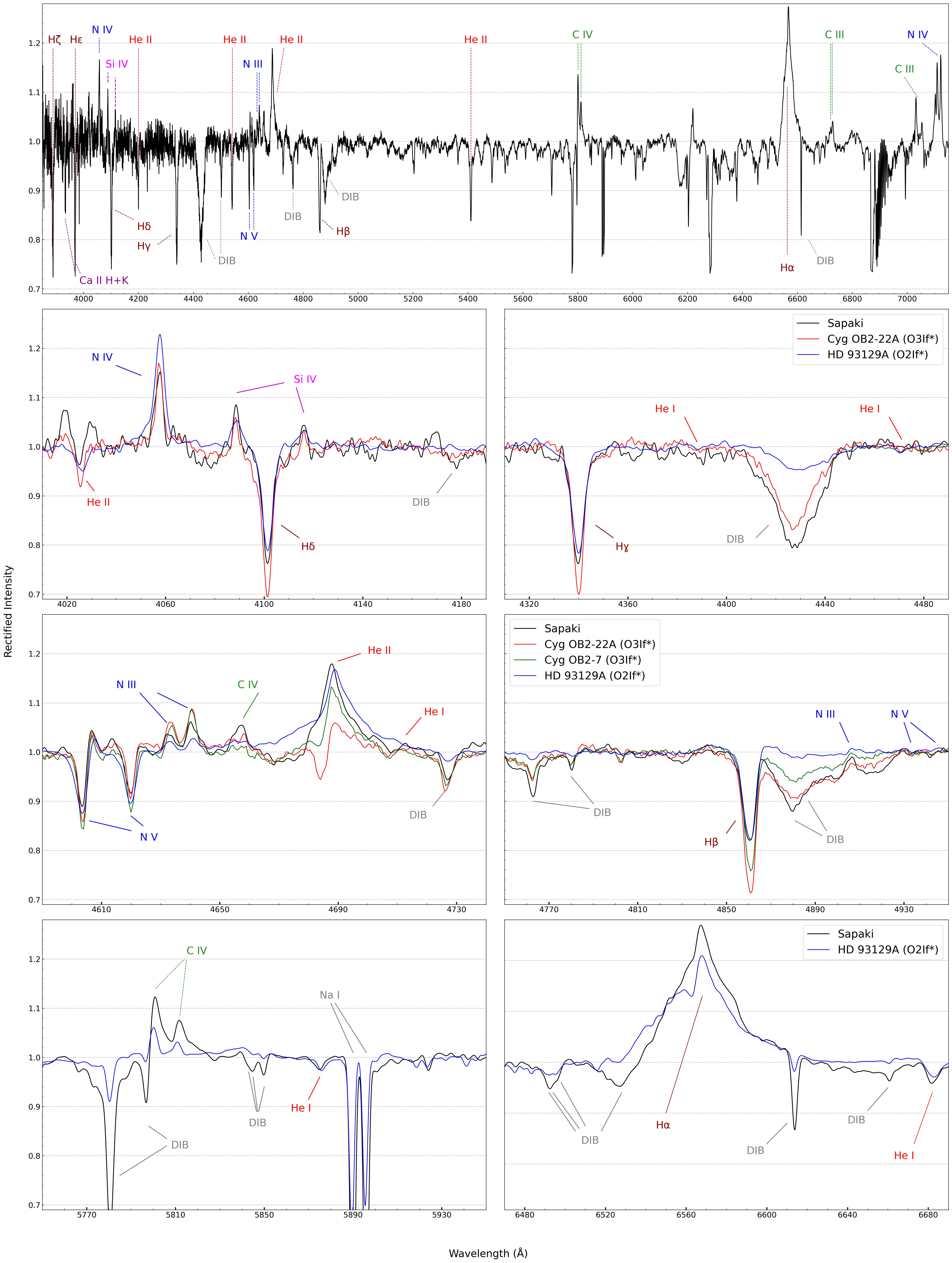}
      \caption{\textit{} MagE optical spectrum of Sapaki in the range 3800$-$7200 \r{A} at a resolution of $R=4100  $, shown at the top. The remaining panels display comparisons between the spectra of Sapaki (black), HD 93129A (O2If*; blue), Cyg OB2-22A (O3If*; red), and Cyg OB2-7 (O3If*; green), in the wavelength ranges of 4010$-$4190, 4310$-$4490, 4590$-$4740, 4760$-$4940, 5750$-$5950, and 6470$-$6690 \r{A}, at a common resolution of $R=2500$. Cyg OB2-7 lacks data for the bluest and reddest regions, whereas Cyg OB2-22A lacks data in the reddest regions. The lines marked with colors are H$\eta$, H$\zeta$, Ca {\small II} H, H$\epsilon$ + Ca {\small II} K, He {\small II} $\lambda$4026, N {\small IV} $\lambda$4058, Si {\small IV} $\lambda\lambda$4089-116, H$\delta$, He {\small II} $\lambda$4200, H$\gamma$, He {\small I} $\lambda$4388, He {\small I} $\lambda$4471, He {\small II} $\lambda$4542, N {\small V} $\lambda\lambda$4604-20, N {\small III} $\lambda\lambda$4634-40, C {\small IV} $\lambda$4658, He {\small II} $\lambda$4686, He {\small I} $\lambda$4713, H$\beta$, N {\small III} $\lambda$4905, N {\small V} $\lambda\lambda$4933-44, He {\small II} $\lambda$5411, C {\small IV} $\lambda\lambda$5801-12, He {\small I} $\lambda$5876, H$\alpha$, He {\small I} $\lambda$6678, C {\small III} $\lambda\lambda$6721-27, C {\small III} $\lambda$7037, and N {\small IV} $\lambda$7116. The DIBs marked are $\lambda$4179, $\lambda$4429, $\lambda$4727, $\lambda\lambda$4762-80, $\lambda\lambda$4881-87, $\lambda\lambda$5780-97, $\lambda\lambda$5843-45-50, $\lambda\lambda$6492-94-98, $\lambda$6527, $\lambda$6613, and $\lambda$6660.}
      \label{fig:Sapaki_Spectra}
    \end{figure*}

For the purposes of our study, all the 16 spectra were sum-combined in order to increase the $S/N$. In addition, the spectra were median-combined in order to identify any residual anomalous signal (i.e., cosmic rays and bad pixels), which were manually masked from the sum-combined spectrum. Thereafter, the sum-combined spectrum was rectified. The final MagE optical spectrum of Sapaki is presented in the top panel of Fig. \ref{fig:Sapaki_Spectra}. 

\section{Results and discussion} \label{sec:results}

    \subsection{Spectral type}

    The top panel of Fig. \ref{fig:Sapaki_Spectra} presents the MagE rectified optical spectrum of Sapaki at a resolution $R=4100$, in which the most important spectral features are marked in color. First of all, the absence of He {\small I} $\lambda4471$,  and the presence of strong He {\small II} $\lambda$4542 and N {\small V} $\lambda\lambda$4604-20 absorption lines, in conjunction with the He {\small II} $\lambda$4686 and N {\small IV} $\lambda$4058 emission line features, are indicative of a very hot O-type star (Walborn \citeyear{W1971}, \citeyear{W1982}; Walborn et al. \citeyear{Wetal2002}). Indeed, the N {\small IV} $\lambda$4058 emission line is stronger (but not by much) than the N {\small III} $\lambda\lambda$4634-41 emission line, and therefore should be classified as an O3 star (Walborn et al. \citeyear{Wetal2002}; Crowther \& Walborn \citeyear{CW2011}; Roman-Lopes et al. \citeyear{RL2016}). Additionally, the strong He {\small II} $\lambda$4686 emission line together with the N {\small III} $\lambda\lambda$4634-41 line (also seen in the emission) has led us to assign it a supergiant status and the f* phenomenon (Sota et al. \citeyear{Setal2011}, \citeyear{Setal2014}; Martins \citeyear{M2018}). Moreover, the H$\beta$ recombination line seen purely in absorption is characteristic of normal O-type star, ruling out the WN classification suggested by Roman-Lopes (\citeyear{RL2011}), as WN stars spectra display H$\beta$ in emission and OIf*/WN stars present H$\beta$ with a P-cygni profile (Crowther \& Walborn \citeyear{CW2011}; Sota et al. \citeyear{Setal2014}; Roman-Lopes et al. \citeyear{RL2016}). Furthermore, we compared its MagE optical spectrum with those of the O2If* prototype HD 93129A and two Galactic O3If* stars: Cyg OB2-22A and Cyg OB2-7. The comparison spectra were retrieved from the Galactic O-Star Spectroscopic Survey (GOSSS; Sota et al. \citeyear{Setal2011}). The spectrum of the O2If* star comes with a resolution $R=4000$, whereas the spectra of the O3If* stars have $R=2500$. Accordingly, the spectrum of HD 93129A and Sapaki were degraded to a common resolution of $R=2500$. In the lower panels of Fig. \ref{fig:Sapaki_Spectra}, we present the comparison between spectra in the following (when available) spectral ranges: 4010$-$4190 \r{A}, 4390$-$4590, 4590$-$4740 \r{A}, 4750$-$4950 \r{A}, 5750$-$5950, and 6470$-$6690 \r{A}. We can notice that the spectrum of Sapaki resembles very well the optical spectra of Cyg OB2-7 and Cyg OB-22A, rather than that of HD 93129A. Therefore, Sapaki should be assigned the spectral type O3If* (Walborn \citeyear{W1982}; Walborn et al. \citeyear{Wetal2002}; Crowther \& Walborn \citeyear{CW2011}; Sota et al \citeyear{Setal2014}; Roman-Lopes et al. \citeyear{RL2016}; Martins \citeyear{M2018}). 

    We emphasize that an accurate spectral classification of Sapaki requires a blue-violet spectrum ($4000-4900$ \r{A}) of moderate resolution ($R=4100$) and moderate signal-to-noise ($S/N=95$ at $4500$ \r{A}). Although its previous spectral type WN6 was discarded, it was first classified as such by Roman-Lopes (\citeyear{RL2011}) with low-resolution ($R=1000$), moderate-signal ($S/N=100$) H- and K-band spectra. Indeed, the spectral classification of early-type stars in the NIR is still difficult to achieve with modest data because of the smaller amount of atomic transitions, easy generation of emission lines due to the amplification of non-LTE effects, and strong influence of stimulated emission on some absorption features (Hanson et al. \citeyear{Hetal2005}; Gray \& Corbally \citeyear{GC2009}; Przybilla \citeyear{P2010}). Therefore, more efforts need to be invested in obtaining high-quality blue-violet spectra of early-type stars despite much longer exposure times. This is because the 4000-4900 \r{A} range holds a high density of lines that are well-behaved with respect to changes in the effective temperature and surface gravity of the stars (Morgan et al. \citeyear{MKK1943}; Morgan \& Keenan \citeyear{MK1973}; Gray \& Corbally \citeyear{GC2009}; Martins \& Plez \citeyear{MP2006}; Martins \citeyear{M2018}).

    \subsection{Interstellar extinction} \label{sec:LOSE}

    In order to estimate the interstellar extinction in the line of sight, we retrieved optical and NIR photometric values of Sapaki from the Transiting Exoplanets Survey Satellite (TESS) and the Two Micron All-Sky Survey (2MASS) databases (see Table \ref{tab:Sapaki_astrophot}). Next, we obtained estimates for the intrinsic colors of O3 supergiants from the work of Martins \& Plez (\citeyear{MP2006}), in which the typical uncertainties are better than $0.10$ mag. On this basis, we computed color excess values $E(B-V)=2.64 \pm 0.15$, $E(V-J)=5.71 \pm 0.16$, $E(V-H)=6.50 \pm 0.16,$ and $E(V-K_{s})=6.96 \pm 0.21$. The optical-to-selective extinction ratio was estimated using  Equations A3-A6 of Fitzpatrick (\citeyear{F1999}), which turned out to provide a mean value of $R_{V} = 2.98\pm0.23$. Finally, by using $R_{V}$ and $E(B-V),$ we determined the total optical extinction to be $A_{V} = 7.87\pm0.75$. This value agrees well with the result obtained by the StarHorse2 survey (Anders et al. \citeyear{Aetal2022}) of $A_{V} = 7.11 \pm 0.13$, although the latter was computed by assuming $d = 3.67 \pm 0.11$ kpc. 
    
    The blueband extinction can be estimated through the color excess $E(B-V)$ and the total optical extinction, $A_{V}$, resulting in $A_{B} = 10.51 \pm 0.95$. At the same time, the extinction in the NIR bands can be evaluated using the color excesses and the intrinsic colors $(V-J)_{0}$, $(V-H)_{0}$, and $(V-K_{s})_{0}$, together with the optical extinction, resulting in values of $A_{J}=2.16 \pm 0.25$, $A_{H}=1.37 \pm 0.25,$ and $A_{K_{s}}=0.91 \pm 0.26$, which translate to mean relative extinctions of $A_{J}/A_{V} = 0.27$, $A_{H}/A_{V} = 0.17,$ and $A_{K_{s}}/A_{V} = 0.12$. Regardless, by using Equations (2a) and (2b) from Cardelli et al. (\citeyear{CCM89})  describing a extinction law in the near-infrared along with the value $R_{V}=2.98$, we were able to compute the mean relative extinctions $0.28$, $0.18,$ and $0.11$ for the $J-$, $H-,$ and $K_{s}$ bands. These results agree well with those derived in the line of sight of Sapaki.

    Furthermore, the value of $R_{V} = 2.98 \pm 0.28$ is consistent with the canonical value $R_{V} = 3.1$ for the diffuse Galactic medium (Tielens \citeyear{T2005}). In order to evaluate the extinction in the M and 8 $\mu$m bands, we adopted the Galactic extinction law from Rieke \& Lebofsky (\citeyear{RL1985}) in the range of 3.0$-$13.0 $\mu$m, which was obtained for $R_{V} = 3.09 \pm 0.03$. Therefore, we assumed $A_{M}/A_{V} = 0.02$ and $A_{8.0}/A_{V} = 0.02$ toward Sapaki. Despite the extinction law approaches to zero at these wavelengths, the values of the distance estimates could be overestimated by at least 7\%, which is significant.

    \subsection{Heliocentric distance}

    \subsubsection{Visual distance modulus} \label{sec:VDM}
    Once the nature of the source is known, we can estimate the heliocentric distance if its absolute magnitude is known beforehand. For example, we can start by assuming typical values for the absolute visual magnitudes of O3I stars $M_{V} = -6.42 \pm 0.10$, taken from the work of Martins \& Plez (\citeyear{MP2006}). This value agrees well with the averaged value of $M_{V} = -6.35 \pm 0.40$ for six O2$-$O3.5If* in Walborn et al. (\citeyear{Wetal2002}). Following, the total optical extinction and $V-$band photometry values can then be applied onto the distance modulus equation corrected by extinction,
    \begin{equation}
        V - M_{V} = 5\log(d) - 5 + A_{V} \label{eq:distance_modulus}
    ,\end{equation}
    and by considering $M_{V} = -6.42 \pm 0.10$, $V = 14.90\pm0.05,$ and $A_{V} = 7.87 \pm 0.93$, we compute a heliocentric distance of $d = 4.89 \pm 1.71$ kpc. The way the distance is calculated magnifies the error coming from photometric quantities and the reddening law, which is a common issue in the context of early-type stars (Walborn \citeyear{Wetal2002}). Despite the uncertainty being as large as 35\%, the obtained quantities are not surprising because the dust accumulation in the Galactic disk seriously compromises the calculation of distances by any method, due to large values of interstellar extinction and the big spread of visual absolute magnitudes for stars earlier than O4 (Walborn et al. \citeyear{Wetal2002}; Gray \& Corbally \citeyear{GC2009}; Wu et al. \citeyear{Wetal2014}).

    \subsubsection{Infrared distance modulus} \label{sec:IDM}
    The use of mid-infrared band photometry may result in lower uncertainties on the estimated distance, as total extinctions in such bands are related to the optical ones: $A_{M}/A_{V} = 0.02$ and $A_{8.0}/A_{V} = 0.02$. Besides, we may provide an estimate for the mass of Sapaki from other Galactic stars earlier than O4. We can cite the following examples (a) Cyg OB2-7 (O3If*) with current mass of $135 M_{\odot}$ (Walborn et al. \citeyear{Wetal2002}), (b) RFS8 (O3.5If*) with a current mass of $\sim70 M_{\odot}$ (Roman-Lopes et al. \citeyear{RL2016}), (c) W49nr1 (O2-3.5If*) with a current mass $>95 M_{\odot}$ (Wu et al. \citeyear{Wetal2014}), or (d) HD 93129 AaAb (O2If* + O2If* + OB?) with current masses well above 60 and 30 $M_{\odot}$, for components Aa and Ab, respectively (Maíz-Apellániz et al. \citeyear{MAetal2017}). 

We then set a rough lower limit of 70 $M_{\odot}$ for the mass of Sapaki, which agrees well with typical masses of O3If* stars found by Martins (\citeyear{Metal2005}). Next, we retrieved 1 Myr isochrones with $A_{V} = 7.87$ and solar metallicity from the work of Bressan et al. (\citeyear{Betal2012}). We found $M_{4.5} = -4.82 \pm 0.15$ mag for $60-80 M_{\odot}$ stars, which already account for the effect of $A_{V}$. Finally, M-band photometry could be estimated from W2-band WISE measurements (Cutri et al. \citeyear{Cetal2012}), as they operate in the same spectral range. We retrieved $W2 = 8.22 \pm 0.02$, which can be plugged into Equation \ref{eq:distance_modulus} to establish a heliocentric distance of $d = 4.41 \pm 0.31$ kpc. If we instead consider a 2 Myr isochrone, we estimate $M_{W2} = -5.77 \pm 0.33,$ which leads to a result of $d = 6.83 \pm 1.05$ kpc. Despite the disparity of both results, the latter is still well below the far-distance scenario proposed by Roman-Lopes (\citeyear{RL2011}), thus we may discard it. As a consequence, the star would be located in the near side of the Scutum-Centaurus arm, with 7 kpc standing as a robust upper limit on the distance to Sapaki.
    
    \subsubsection{Parallaxes from Gaia}

    The Gaia Data Release 3 (GDR3) catalog contains improved astrometric and photometric measurements collected over the course of 34 months for almost two billion objects accross the sky (Gaia Collaboration \citeyear{Gaia2022}). However, the published parallax values are biased because of imperfections in the instruments and data processing methods (Lindegren et al. \citeyear{Lietal2021}), thus they have to be corrected first in order to estimate the heliocentric distance by inverting the parallax as follows,
    \begin{align}
        d &= 1 \; / \; \bar{\omega_{c}} \label{eq:distance_parallax}
    \end{align}
    where $\bar{\omega_{c}} = \bar{\omega} - Z_{\text{DR3}}$ is the corrected parallax, $\bar{\omega}$ is the published parallax and $Z_{\text{DR3}}$\footnote{In Maíz-Apellániz et al. (\citeyear{MAetal2022}), the parallax zero point is calculated for the Early Gaia Data Release 3 (EDR3), however for Sapaki we have $Z_{\text{EDR3}} = Z_{\text{DR3}}$ since none of its parameters were updated} is the parallax zero point for GDR3. The value $Z_{\text{DR3}}$ depends on the $G$-band photometry, the effective wavenumber, the ecliptic latitude and the quality of the astrometric solution of the source (Lindegren et al. \citeyear{Lietal2021}). In Maíz-Apellániz et al. (\citeyear{MAetal2022}) a new way of determining $Z_{\text{DR3}}$ was presented, whereby we take into account the distribution of blue and red stars, either in the Milky Way or the Magellanic Clouds. We may estimate the zero point by using the routine \texttt{ZPEDR3} of Table A.2 in Maíz-Apellániz et al. (\citeyear{MAetal2022}), which leads us to calculate $Z_{\text{DR3}} = -0.056$ mas. 

We found the corrected parallax to be $\bar{\omega_{c}} = 0.247$ mas. Thus, by applying Equation \ref{eq:distance_parallax}, we calculated $d = 4.11$ kpc. However, the real uncertainty is greatly underestimated because of crowding and extinction effects, together with the limited magnitude of the instrument and quality of the astrometric solution (Lindegren et al. \citeyear{Lietal2021}). The external uncertainty, $\sigma_{\text{ext}}$, depends on the published parallax uncertainty, the $G$-band photometry, the RUWE value, and the quality of the astrometric solution of the star. This can be estimated by using the routine \texttt{SPICOR} of Table A.1 in Maíz-Apellániz et al. (\citeyear{MAetal2022}), which then leads us to calculate $\sigma_\text{ext} = 0.128$ mas, a value that is a factor of 3 greater than the published parallax uncertainty (see Table \ref{tab:Sapaki_astrophot}). Next, by propagating the external uncertainty for the distance, we found $\sigma_{d} = 2.16$ kpc. Therefore, from a Gaia-based analysis, the distance of Sapaki would be $d = 4.11 \pm 2.16$ kpc, which is even more inaccurate than when the visual distance modulus is used.

    Aside from our result, we may take a look at the heliocentric distances computed by Bailer-Jones et al. (\citeyear{BJetal2021}), who provided two possible distance estimates for objects in GDR3 based on isotropic Bayesian priors to compute a geometric distance using only the parallax and its uncertainty, as well as a photogeometric distance that includes the (extincted) color and magnitude of the star. The latter result is less accurate at lower Galactic latitudes, therefore, for Sapaki, the geometric distance $d = 4.20^{+0.88}_{-0.55}$ kpc seems more adequate. Although the formalism of Bailer-Jones et al. (\citeyear{BJetal2021}) is primarily based on the spatial distribution of late-type stars, which is not strictly applicable to OB stars, the resultant distance is quite similar to that obtained by applying the corrections listed in Maíz-Apellániz et al. (\citeyear{MAetal2022}). Given the fact we obtained a distance on the order of 4 kpc, we may set this value as a robust lower limit for the heliocentric distance of Sapaki.

    \subsection{Radial velocity}

    Based on the MagE optical spectrum of Sapaki (see Fig. \ref{fig:Sapaki_Spectra}), we were able to measure the center of several spectral lines, allowing us to estimate the radial velocity of the star. In Table \ref{tab:Sapaki_vr}, we present measurements of the estimated center $\lambda_{\text{obs}}$ for the main observed spectral lines. Every line was measured at least ten times by fitting both Gaussian and Voigt profiles through \texttt{splot} task on IRAF. Typical errors in $\lambda_{\text{obs}}$ are $\pm$0.1 \r{A} and lead to a systematic error of $\pm$5 km/s on individual measurements. The resultant radial velocity for Sapaki is $V_{r} = -34.2 \pm 15.6$ km/s when all uncertainties are taken into account (see Table \ref{tab:Sapaki_vr}). The resultant value is consistent with the radial velocity measured by Caswell \& Haynes (\citeyear{CH1987}) for the H {\small II} region G321.1-0.55.

    \begin{table}[htbp]
        \centering
        \begin{tabular}{c|c|c|c|c}
            \hline
            Feature & $\lambda_{0}$ (\r{A}) & $\lambda_{\text{obs}}$ (\r{A}) & $\Delta\lambda$ (\r{A}) & $V_{r}$ (km/s) \\
            \hline
            N {\small IV} & 4057.76 & 4057.34 & -0.42 & -31.0 \\
            Si {\small IV} & 4088.85 & 4088.53 & -0.32 & -23.5 \\
            H$\delta$ & 4101.73 & 4101.35 & -0.38 & -28.1 \\
            He {\small II} & 4199.86 & 4199.32 & -0.54 & -38.5 \\
            H$\gamma$ & 4340.47 & 4339.73 & -0.74 & -51.2 \\
            He {\small II} & 4541.61 & 4541.11 & -0.50 & -33.0 \\
            N {\small V} & 4603.73 & 4603.32 & -0.41 & -26.7 \\
            N {\small V} & 4619.98 & 4619.63 & -0.35 & -22.7 \\
            H$\beta$ & 4861.29 & 4860.50 & -0.79 & -48.5 \\
            He {\small II} & 5411.52 & 5410.83 & -0.69 & -38.2 \\
            \hline
        \end{tabular}
        \caption{Line-center measurements of several absorption/emission present in the atmosphere of Sapaki. $\lambda_{0}$ is the air wavelength, $\lambda_{\text{obs}}$ is the measured wavelength with \texttt{splot} routine in IRAF, $\Delta\lambda = \lambda_{\text{obs}} - \lambda_{0}$, and $V_{r} = c \Delta\lambda/\lambda_{\lambda_{0}}$, where $c = 2.99 \times 10^{5}$ km/s.}
        \label{tab:Sapaki_vr}
    \end{table}
    
    \subsection{Examining a plausible case of isolated star formation}

    The question of whether a high-mass star can be born in isolation is still a matter of debate, although it has been found that the majority of field massive stars are part of the Galactic runaway stellar population (de Wit et al. \citeyear{dWetal2005}; Perets \& Šubr \citeyear{PS2012}; Dorigo-Jones et al. \citeyear{DJetal2020}). On the other hand, if a massive star is immersed in a H {\small II} region without a bow shock, then the likelihood of being a true massive star born in the field rises (Gvaramadze et al. \citeyear{Getal2012}; Stephens et al. \citeyear{Setal2017}; Vargas-Salazar et al. \citeyear{VSetal2020}). Finally, if the local IMF is the result of a more stochastic process, there should not be any restriction on the mass of the second most massive member and we would expect a number of low-mass stars clustered around the main source, with few B- and A-type stars (Kalari et al. \citeyear{Ketal2019}).

    Focusing back to Sapaki, in Fig. \ref{fig:Sapaki_image}, we can see that the star is close to the center of a well-developed, nearly circular cavity, which is not expected in the case of a runaway star. However, the non-detection of bow shocks is not a sufficient condition for ruling out a runaway status because about 80\% of runaway OB stars do not present observable bow shocks due to their weak emission (below certain detection limit) or because they simply do not form since the stellar space velocity is not as high as the speed of sound in the local ISM (Gvaramadze et al. \citeyear{Getal2012}).
    
\begin{figure}[]
      \centering
      \includegraphics[width=3in]{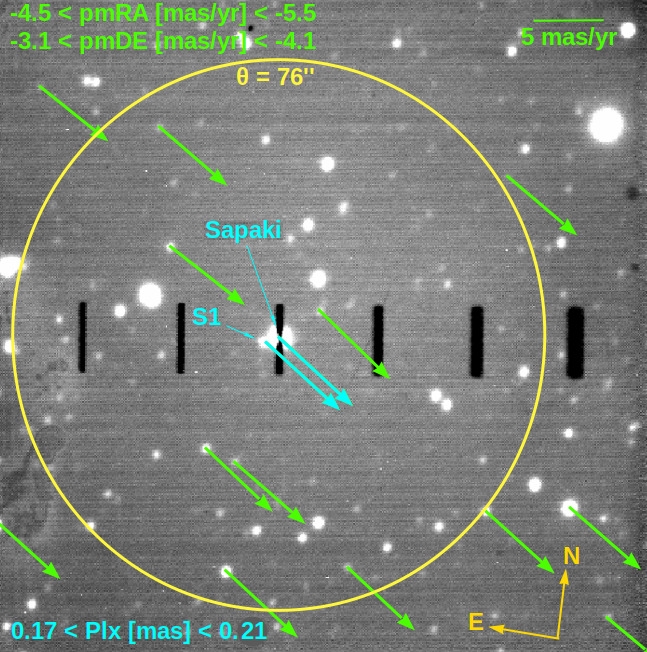}
      \caption{ $86\times82$ arcsec$^{2}$ optical image taken with the slit viewing camera of MagE. Sapaki and S1 are labeled in cyan at the center of the 1.0$''$ slit. The yellow circle has an angular diameter $\theta = 76''$. We can see that only S1 has similar proper motion and parallax measurements to those of Sapaki.}
      \label{fig:Sapaki_acq}
    \end{figure}

    An alternative analysis for its runaway status can be done by using high-quality proper motions published in the GDR3. We start by transforming proper motions in RA ($\mu_{\alpha}$) and declination ($\mu_{\delta}$) into their equivalents in Galactic latitude ($\mu_{b}$) and longitude ($\mu_{l}$), which were computed using the routine \texttt{SkyCoord} from \texttt{astropy.coordinates} package within Python. We obtained $\mu_{l} = -6.17 \pm 0.04$ mas/yr and $\mu_{b} = -0.44 \pm 0.05$ mas/yr. Then, we compared the results with the Galactic distribution in ($\mu_{b}$, $l$) and ($\mu_{l}$, $l$) space for O-type stars (see Figs. 1 and 2 in Maíz-Apellániz et al. \citeyear{MAetal2018}). For a quantitative comparison, we have to calculate the normalized difference between the catalogued proper motions and those from the Galactic distribution of O-type stars, which is
    \begin{equation}
        \Delta = \sqrt{ \bigg( \frac{\mu_{b,\text{rel}}}{\sigma_{\mu_{b}}} \bigg)^{2} + \bigg( \frac{\mu_{l,\text{rel}}}{\sigma_{\mu_{l}}} \bigg)^{2} } \label{eq:Delta_runaway}
    ,\end{equation}
    where $\mu_{l,\text{rel}} = \mu_{l} - \bar{\mu_{l}}$ and $\mu_{l,\text{rel}} = \mu_{b} - \bar{\mu_{b}}$. The value $\bar{\mu_{l}} = a_{0} + a_{1}\cos(l) + a_{2}\cos(2l)$ must be calculated for the Galactic longitude of Sapaki, which turned out to be $\bar{\mu_{l}} = -3.97$ mas/yr. The values of $a_{0}$, $a_{1}$, $a_{2}$, $\bar{\mu_{b}}$, $\sigma_{\mu_{l}}$, and $\sigma_{\mu_{b}}$ can be retrieved from Table 8 of Maíz-Apellániz et al. (\citeyear{MAetal2018}). Once the previous were plugged into Equation \ref{eq:Delta_runaway}, we found $\Delta = 1.3$. The derived value is below the empirically-established cut of $3.5$ for detecting runaway stars (Tetzlaff et al. \citeyear{Tetal2011}; Maíz-Apellániz et al. \citeyear{MAetal2018}). Therefore, from a 2D kinematic analysis, we find that Sapaki is not a runaway star. Moreover, in Fig. \ref{fig:Sapaki_acq}, we notice that its proper motion is similar to that of the nearby population, which lends support to the idea of its being formed in situ, rather than being a runaway. Nevertheless, a full 3D analysis that includes radial velocity measurements of the stars surrounding Sapaki is required to completely rule out this possibility, because its radial velocity ($V_{r} = -34.2 \pm 15.6$ km/s) might remain the same for the field population. (de Wit et al. \citeyear{dWetal2005}; Tetzlaff et al. \citeyear{Tetal2011}; Gvaramadze et al. \citeyear{Getal2012}; Kalari et al. \citeyear{Ketal2019}; Sana et al. \citeyear{Setal2022}).

    From a close look of Fig. \ref{fig:Sapaki_image}, we do not perceive any luminous member within a circle of angular diameter $\theta = 76''$ around Sapaki. Additionally, in Fig. \ref{fig:Sapaki_acq}, we present an optical image of the immediate vicinity around Sapaki retrieved from the slit viewing camera of MagE, in which we did not detect any significant local clustering and which gives support to our assumption of a truly isolated star. On the other hand, in Fig. \ref{fig:Sapaki_acq}, we have plotted proper motions from GDR3 in the range $-4.5<\mu_{\alpha}/\text{(mas/yr)}<-5.5$ and $-3.1<\mu_{\delta}/\text{(mas/yr)}<-4.1$, from which we detected 14 sources with proper motions similar to that of Sapaki. Even so, only one source (labeled as S1 and marked in cyan in Fig. \ref{fig:Sapaki_image}) also has  a parallax value in the range $0.17<\bar{\omega}/\text{(mas)}<0.21$. Such range of parallax values correspond to a depth of $\sim1$ kpc. It is remarkable that S1 has a \texttt{ruwe} value of $1.02$ that indicates well-behaved astrometric solutions. By following the same parallax-correction procedure from Maíz-Apellániz et al. (\citeyear{MAetal2022}), as for Sapaki, we found $\sigma_{\text{ext}} = 0.185$ mas and $Z_{\text{GDR3}} = -0.055$ mas, which translates to a distance of $d = 4.35 \pm 3.49$ kpc, whereas from Bailer-Jones et al. (\citeyear{BJetal2021}) we retrieved a geometric distance of $d = 4.95^{+1.61}_{-1.49}$ kpc. Although both distances agree well with the absolute value of Sapaki, the uncertainties are huge and therefore not recommended for use in stellar parameter calculation. On top of that, given the very slight difference in terms of the measured proper motions, corrected parallaxes (distances), and angular separation in the sky, Sapaki and S1 may possibly be part of the same system. If so, we may discard the runaway status with respect to Sapaki because multiple stellar systems argue against the observations of many runaway OB stars being single systems (de Wit et al. \citeyear{dWetal2005}; Sana et al. \citeyear{Setal2014}; Oh \& Kroupa \citeyear{OK2016}). Considering the exposed above, Sapaki appears to be a strong candidate for isolated massive stellar formation in the Milky Way that deserves to be studied with other methods, such as multi-epoch high-resolution spectroscopy, to obtain accurate radial velocity values. Also, deep high-resolution imaging could reveal any stellar cluster around Sapaki that may be still hidden.

\section{Summary} \label{sec:conclusion}

    In this study, we report the identification of a new Galactic O3If* star, classified as such on the basis of the resemblance of  its characteristics to those of the O3If* class. We computed a total-to-selective extinction parameter $R_{V} = 2.98\pm0.23$, which allowed us to derive a total optical extinction of $A_{V} = 7.87\pm0.75$, a blue extinction of $A_{B} = 10.51\pm0.95,$ and mean relative extinctions of $A_{J}/A_{V} = 0.27$, $A_{H}/A_{V} = 0.17$, $A_{K_{s}}/A_{V} = 0.12$, $A_{M}/A_{V} = 0.02,$ and $A_{8.0}/A_{V} = 0.02$. These results  are consistent with the standard Galactic extinction law of Cardelli et al. (\citeyear{CCM89}). We measured its radial velocity from several spectral lines, obtaining $V_{r} = -34.2 \pm 15.6$ km/s. From several approaches, we were able to estimate that the heliocentric distance of Sapaki is probably in the range of $4 - 7$ kpc. The large scatter of the result arises due to enormous extinction values in the Galactic plane that bias parallax and photometric measurements, in addition to the large scatter in the determination of absolute magnitudes in the most massive stars. Nevertheless, the derived values agree well with the near-distance scenario proposed by Roman-Lopes (\citeyear{RL2011}), which places Sapaki in the near side of the Scutum-Centaurus arm.

    Finally, we present evidence suggesting that Sapaki is a strong candidate of isolated massive star formation. As first noticed by Roman-Lopes (\citeyear{RL2011}), we can see that Sapaki appears as the only point-like source in 8.0 $\mu$m images and it is very close to the center of a nearly-circular cavity delineated by hot dust emission. Moreover, we may discard a runaway status for Sapaki if we consider solely its 2D kinematics,which is consistent with the velocity distribution of Galactic O-type stars. Furthermore, by applying constraints to GDR3 proper motions and parallaxes, we detected only one source (labeled as S1 on Fig. \ref{fig:Sapaki_acq}) with compatible astrometric measurements to those obtained for Sapaki. In order to rule out the possibility of Sapaki being a runaway star, our team is performing a full 3D kinematic analysis which includes NIR radial velocity measurements of the field population around the coordinates of Sapaki. Consequently, Sapaki would be an excellent candidate for demonstrating isolated massive star formation under Milky Way conditions, which would make it a milestone for theoretical studies of massive star formation.
    
\begin{acknowledgements}
      The authors would like to thank the anonymous referee for the excellent comments and suggestions which greatly contributed to improve the original manuscript. M.S.Z.P. is grateful for the financial support provided by Becas ANID-Chile, and for the constant guidance of A.R.L. and D.S. This work includes data gathered with the 6.5-meter Magellan Baade Telescope located at Las Campanas Observatory, Chile. We thank the staff for the efficient support provided during the MagE observing runs. IRAF (Tody \citeyear{T1986}; Tody \citeyear{T1993}) is distributed by the National Optical Astronomy Observatory, which is operated by the Association of Universities for Research in Astronomy (AURA) under a cooperative agreement with the National Science Foundation. CarPy (Kelson et al. \citeyear{Ketal2000}; Kelson \citeyear{K2003}) is distributed by the Carnegie Observatories, which are operated by the Carnegie Institution for Science. This research makes use of data products from the \textit{2MASS}, which is a joint project of the University of Massachusetts and the Infrared Processing and Analysis Center/California Institute of Technology. This research makes use of data products from the space mission Gaia, which is a joint project from the European Space Agency and the Gaia Data Processing and Analysis Consortium. This publication makes use of data products from \textit{WISE}, which is a joint project of the University of California, Los Angeles, and the Jet Propulsion Laboratory/California Institute of Technology, funded by the National Aeronautics and Space Administration. This research has made use of the SIMBAD database, the VizieR database and ALADIN Interactive Sky Atlas (Bonnarel et al. \citeyear{Betal2000}) operated at CDS, Strasbourg, France.
\end{acknowledgements}



\begin{thebibliography}{}

    \bibitem[Anders et al.(2022)]{Aetal2022} Anders, F., Khalatyan, A., Queiroz, A. B. A., et al. 2022, \href{https://doi.org/10.1051/0004-6361/202142369}{\color{blue}A\&A}\href{https://ui.adsabs.harvard.edu/abs/2022A\%26A...658A..91A/}{, 658, 91} 

    \bibitem[Anderson et al.(2014)]{Aetal2014} Anderson, L. D., Bania, T. M., Danas, S. B., et al. 2014, \href{https://doi.org/10.1088/0067-0049/212/1/1}{\color{blue}ApJS}\href{https://ui.adsabs.harvard.edu/abs/2014ApJS..212....1A/}{, 212, 1} 

    \bibitem[Bailer-Jones et al.(2021)]{BJetal2021} Bailer-Jones, C., Rybizki, J., Fouesneau, M., et al. 2021, \href{https://doi.org/10.3847/1538-3881/abd806}{\color{blue}AJ}\href{https://ui.adsabs.harvard.edu/abs/2021yCat.1352....0B/}{, 161, 147} 

    \bibitem[Benjamin et al.(2003)]{Betal2003} Benjamin, R. A., Churchwell, E., Babler, B. L., et al. 2003, \href{https://doi.org/10.1086/376696}{\color{blue}PASP}\href{https://ui.adsabs.harvard.edu/abs/2003PASP..115..953B/}{, 115, 953} 

    \bibitem[Bonnarel et al.(2000)]{Betal2000} Bonnarel, F., Fernique, P., Bienaymé, O., et al. 2000, \href{https://doi.org/10.1051/aas:2000331}{\color{blue}A\&AS}\href{https://ui.adsabs.harvard.edu/abs/2000A\%26AS..143...33B/}{, 143, 33} 

    \bibitem[Bressan et al.(2012)]{Betal2012} Bressan, A., Marigo, P., Girardi, L., et al. 2012, \href{https://doi.org/10.1111/j.1365-2966.2012.21948.x}{\color{blue}MNRAS}\href{https://ui.adsabs.harvard.edu/abs/2012MNRAS.427..127B/}{, 427, 127} 

    \bibitem[Caswell \& Haynes(1987)]{CH1987} Caswell, J. L., \& Haynes, R. F., 1987, \href{https://adsabs.harvard.edu/pdf/1987A\%26A...171..261C}{\color{blue}A\&A}\href{https://ui.adsabs.harvard.edu/abs/1987A\%26A...171..261C/}{, 171, 261} 

    \bibitem[Clayton, Cardelli \& Clayton(1989)]{CCM89} Cardelli, J. A., Clayton, G. C., and Mathis, J. S., 1989, \href{https://doi.org/10.1086/167900 }{\color{blue}ApJ}\href{https://ui.adsabs.harvard.edu/abs/1989ApJ...345..245C/}{, 345, 245} 

    \bibitem[Crowther \& Walborn(2011)]{CW2011} Crowther, P., \& Walborn, N., 2011, \href{https://doi.org/10.1111/j.1365-2966.2011.19129.x}{\color{blue}MNRAS}\href{https://ui.adsabs.harvard.edu/abs/2011MNRAS.416.1311C/}{, 416, 1311} 

    \bibitem[Cutri et al.(2012)]{Cetal2012} Cutri, P., et al. 2012, \href{https://cdsarc.cds.unistra.fr/viz-bin/cat/II/311#/prov}{\color{blue}VizieR}\href{https://ui.adsabs.harvard.edu/abs/2012yCat.2311....0C/}{, 2311} 

    \bibitem[de Wit et al.(2005)]{dWetal2005} de Wit, W. J., Testi, L., Palla, F., \& Zinnecker, H., 2005, \href{https://doi.org/10.1051/0004-6361:20042489}{\color{blue}A\&A}\href{https://ui.adsabs.harvard.edu/abs/2005A\%26A...437..247D/}{, 437, 247} 

    \bibitem[Dorigo-Jones et al.(2020)]{DJetal2020} Dorigo Jones, J., Oey, M. S., Paggeot, K., et al. 2020, \href{https://doi.org/10.3847/1538-4357/abbc6b}{\color{blue}ApJ}\href{https://ui.adsabs.harvard.edu/abs/2020ApJ...903...43D}{, 903, 43} 

    \bibitem[Fierlinger et al.(2016)]{Fetal2016} Fierlinger, K. M., Burkert, A., Ntormousi, E., et al. 2016, \href{https://doi.org/10.1093/mnras/stv2699}{\color{blue}MNRAS}\href{https://ui.adsabs.harvard.edu/abs/2016MNRAS.456..710F/}{, 456, 710} 

    \bibitem[Fitzpatrick(1999)]{F1999} Fitzpatrick, E., 1999, \href{https://doi.org/10.1086/316293}{\color{blue}PASP}\href{https://ui.adsabs.harvard.edu/abs/1999PASP..111...63F/}{, 111, 63} 

    \bibitem[Gaia Collaboration(2022)]{Gaia2022} Gaia Collaboration: Vallerani, A., Brown, A. G. A., Prusti, T., et al. 2022, \href{https://doi.org/10.1051/0004-6361/202243940}{\color{blue}A\&A}, \href{https://ui.adsabs.harvard.edu/abs/2022arXiv220800211G/}{Forthcoming article} 

    \bibitem[Gray \& Corbally(2009)]{GC2009} Gray, R. O., \& Corbally, C. J., 2009, Stellar Spectral Classification. Princeton University Press, New Jersey.  

    \bibitem[Gvaramadze et al.(2011)]{Getal2011} Gvaramadze, V. V., Pflamm-Altenburg, J., \& Kroupa, P., 2011, \href{http://dx.doi.org/10.1051/0004-6361/201015656}{\color{blue}A\&A}\href{https://ui.adsabs.harvard.edu/abs/2011A\%26A...525A..17G/}{, 525, 17} 

    \bibitem[Gvaramadze et al.(2012)]{Getal2012} Gvaramadze, V. V., Weidner, C., Kroupa, P., \& Pflamm-Altenburg, J., 2012, \href{https://doi.org/10.1111/j.1365-2966.2012.21452.x}{\color{blue}MNRAS}\href{https://ui.adsabs.harvard.edu/abs/2012MNRAS.424.3037G/}{, 424, 3037} 

    \bibitem[Hanson et al.(2005)]{Hetal2005} Hanson, M. M., Puls, J., \& Repolust, T., 2005, \href{https://doi.org/10.1017/S1743921305004771}{\color{blue}IAUS}\href{https://ui.adsabs.harvard.edu/abs/2005IAUS..227..376H/}{, 227, 276} 

    \bibitem[Kalari et al.(2019)]{Ketal2019} Kalari, V., Vink, J., Wit, W., et al. 2019, \href{https://doi.org/10.1051/0004-6361/201935107}{\color{blue}A\&A}\href{https://ui.adsabs.harvard.edu/abs/2019A\%26A...625L...2K/}{, 625, 2} 

    \bibitem[Kelson(2003)]{K2003} Kelson, D.D., 2003, \href{https://doi.org/10.1086/375502}{\color{blue}PASP}\href{https://ui.adsabs.harvard.edu/abs/2003PASP..115..688K/}{, 115, 688} 

    \bibitem[Kelson et al.(2000)]{Ketal2000} Kelson, D.D., Illingworth, G.D., van Dokkum, P.G., \& Franx, M., 2000, \href{https://doi.org/10.1086/308445}{\color{blue}ApJ}\href{https://ui.adsabs.harvard.edu/abs/2000ApJ...531..159K/}{, 531, 159} 

    \bibitem[Krause et al.(2013)]{Ketal2013} Krause, M., Fierlinger, K., Diehl, R., et al. 2013, \href{http://doi.org/10.1051/0004-6361/201220060}{\color{blue}A\&A}\href{https://ui.adsabs.harvard.edu/abs/2013A\%26A...550A..49K/}{, 550, 49} 

    \bibitem[Lennon et al.(2018)]{Letal2018} Lennon, D. J., Evans, C. J., van der Marel, R. P., et al. 2018, \href{https://doi.org/10.1051/0004-6361/201833465}{\color{blue}A\&A}\href{https://ui.adsabs.harvard.edu/abs/2018A\%26A...619A..78L/}{, 619, A78} 

    \bibitem[Lindegren et al.(2021)]{Lietal2021} Lindegren, L., Bastian,  U., Biermann, M., et al. 2021, \href{https://doi.org/10.1051/0004-6361/202039709}{\color{blue}A\&A}\href{https://ui.adsabs.harvard.edu/abs/2021A\%26A...649A...4L/}{, 649, A4} 

    \bibitem[Maíz-Apellániz et al.(2017)]{MAetal2017} Maíz-Apellániz, J., Sana, H., Barbá, R. H., et al. 2017, \href{https://doi.org/10.1093/mnras/stw2618}{\color{blue}MNRAS}\href{https://ui.adsabs.harvard.edu/abs/2017MNRAS.464.3561M/}{, 464, 3561} 

    \bibitem[Maíz-Apellániz et al.(2018)]{MAetal2018} Maíz-Apellániz, J., Pantaleoni González, M., Barbá, R. H., et al. 2018, \href{https://doi.org/10.1051/0004-6361/201832787}{\color{blue}A\&A}\href{https://ui.adsabs.harvard.edu/abs/2018A\%26A...616A.149M/}{, 616, 149} 

    \bibitem[Maíz-Apellániz et al.(2022)]{MAetal2022} Maíz-Apellániz, J., Pantaleoni González, M., Barbá, R. H., \& Weiler, M., 2022, \href{https://doi.org/10.1051/0004-6361/202142366}{\color{blue}A\&A}\href{https://ui.adsabs.harvard.edu/abs/2022A&A...657A..72M}{, 657, 72} 

    \bibitem[Martins(2018)]{M2018} Martins, F., 2018, \href{https://doi.org/10.1051/0004-6361/201833050}{\color{blue}A\&A}\href{https://ui.adsabs.harvard.edu/abs/2018A\%26A...616A.135M/}{, 616, 135} 

    \bibitem[Martins et al.(2005)]{Metal2005} Martins, F., Schaerer, D., \& Hillier, D. J., 2005, \href{https://doi.org/10.1051/0004-6361:20042386}{\color{blue}A\&A}\href{https://ui.adsabs.harvard.edu/abs/2005A\%26A...436.1049M/}{, 436, 1049}

    \bibitem[Martins \& Plez(2006)]{MP2006} Martins, F., \& Plez, B., 2006, \href{https://doi.org/10.1051/0004-6361:20065753}{\color{blue}A\&A}\href{https://ui.adsabs.harvard.edu/abs/2006A\%26A...457..637M/}{, 457, 637} 

    \bibitem[McKee \& Ostriker(2007)]{MO2007} McKee, C. F., \& Ostriker, E. C., 2007, \href{https://doi.org/10.1146/annurev.astro.45.051806.110602}{\color{blue}ARA\&A}\href{https://ui.adsabs.harvard.edu/abs/2007ARA&A..45..565M/}{, 45, 565} 
    
    \bibitem[Morgan \& Keenan(1973)]{MK1973} Morgan, W. W., \& Keenan, P. C., 1973, \href{https://doi.org/10.1146/annurev.aa.11.090173.000333}{\color{blue}ARA\&A}\href{https://ui.adsabs.harvard.edu/abs/1973ARA\%26A..11...29M/}{, 11, 26} 

    \bibitem[Morgan et al.(1943)]{MKK1943} Morgan, W. W., Keenan, P. C., \& Kellman, E., 1943, An atlas of stellar spectra, with an outline of spectral classification. The University of Chicago Press, Chicago. 

    \bibitem[Oh \& Kroupa(2016)]{OK2016} Oh, S., \& Kroupa, P., 2016, \href{https://doi.org/10.1051/0004-6361/201628233}{\color{blue}A\&A}\href{https://ui.adsabs.harvard.edu/abs/2016A\%26A...590A.107O/}{, 590, 107} 

    \bibitem[Paegert et al.(2021)]{Petal2021} Paegert, M., Stassun, K. G., Collins, K. A., et al. 2021, \href{https://doi.org/10.48550/arXiv.2108.04778}{\color{blue}arXiv}\href{https://ui.adsabs.harvard.edu/abs/2021arXiv210804778P/}{:2108.04778} 

    \bibitem[Perets \& Šubr(2012)]{PS2012} Perets, H. B., \& Šubr, L., 2012, \href{https://doi.org/10.1088/0004-637X/751/2/133}{\color{blue}ApJ}\href{https://ui.adsabs.harvard.edu/abs/2012ApJ...751..133P/}{, 751, 133} 

    \bibitem[Przybilla(2010)]{P2010} Przybilla, N., 2010, \href{https://doi.org/10.1051/eas/1043015}{\color{blue}EAS}\href{https://ui.adsabs.harvard.edu/abs/2010EAS....43..199P/}{, 43, 199} 

    \bibitem[Rieke \& Lebofsky(1985)]{RL1985} Rieke, G. H., \& Lebofsky, M. J., 1985, \href{https://doi.org/10.1086/162827}{\color{blue}ApJ}\href{https://ui.adsabs.harvard.edu/abs/1985ApJ...288..618R/}{, 288, 618} 

    \bibitem[Roman-Lopes(2011)]{RL2011} Roman-Lopes, A., 2011, \href{https://doi.org/10.1111/j.1365-2966.2010.17431.x}{\color{blue}MNRAS}\href{https://ui.adsabs.harvard.edu/abs/2011MNRAS.410..161R/}{, 410, 161}  

    \bibitem[Roman-Lopes(2013)]{RL2013} Roman-Lopes, A., 2013, \href{https://doi.org/10.1093/mnrasl/slt100}{\color{blue}MNRAS}\href{https://ui.adsabs.harvard.edu/abs/2013MNRAS.435L..73R/}{, 435, 73} 

    \bibitem[Roman-Lopes(2011)]{RLetal2011} Roman-Lopes, A., Barba, R. H., \& Morrell, N. I., 2011, \href{https://doi.org/10.1111/j.1365-2966.2011.19062.x}{\color{blue}MNRAS}\href{https://ui.adsabs.harvard.edu/abs/2011MNRAS.416..501R/abstract}{, 416, 501} 

    \bibitem[Roman-Lopes et al.(2016)]{RL2016} Roman-Lopes, A., Franco, G. A. A., \& Sanmartim, D., 2016, \href{https://doi.org/10.3847/0004-637X/823/2/96}{\color{blue}ApJ}\href{https://ui.adsabs.harvard.edu/abs/2016ApJ...823...96R/}{, 823, 96} 

    \bibitem[Sana et al.(2013)]{Setal2013} Sana, H., de Koter, A., de Mink, E., et al. 2013, \href{https://doi.org/10.1051/0004-6361/201219621}{\color{blue}A\&A}\href{https://ui.adsabs.harvard.edu/#abs/2013A\%26A...550A.107S}{, 550, 107} 

    \bibitem[Sana et al.(2022)]{Setal2022} Sana, H., Ramírez-Agudelo, O. H., Brunet-Hénault, V., et al. 2022, \href{https://doi.org/10.1051/0004-6361/202244677}{\color{blue}A\&A}\href{https://ui.adsabs.harvard.edu/abs/2022A\%26A...668L...5S/}{, 668, 5} 

    \bibitem[Skrutskie et al.(2006)]{Setal2006} Skrutskie, M. F., Cutri, R. M., Stiening, R., et al. 2006, \href{https://doi.org/10.1086/498708}{\color{blue}AJ}\href{https://ui.adsabs.harvard.edu/abs/2006AJ....131.1163S/}{, 131, 1163} 

    \bibitem[Smith(2006)]{S2006} Smith, N, 2006, \href{https://doi.org/10.1111/j.1365-2966.2006.10007.x}{\color{blue}MNRAS}\href{https://ui.adsabs.harvard.edu/abs/2006MNRAS.367..763S/}{, 367, 763} 

    \bibitem[Sota et al.(2011)]{Setal2011} Sota, A., Maíz-Apellániz, J., Walborn, N. R., et al. 2011, \href{https://doi.org/10.1088/0067-0049/193/2/24}{\color{blue}ApJS}\href{https://ui.adsabs.harvard.edu/abs/2011ApJS..193...24S/}{, 193, 24} 

    \bibitem[Sota et al.(2014)]{Setal2014} Sota, A., Maíz-Apellániz, J., Morrell, N. I., et al. 2014, \href{https://doi.org/10.1088/0067-0049/211/1/10}{\color{blue}ApJS}\href{https://ui.adsabs.harvard.edu/abs/2014ApJS..211...10S/}{, 211, 10} 

    \bibitem[Steinberg et al(2003)]{Setal2003} Steinberg, A., Hoffmann, T. L., \& Pauldrach, W. A., 2003, \href{htpps://doi.org/10.1086/379506}{\color{blue}ApJ}\href{https://ui.adsabs.harvard.edu/abs/2003ApJ...599.1333S/}{, 599, 1333} 

    \bibitem[Stephens et al.(2017)]{Setal2017} Stephens, I. W., Gouliermis, D., Leslie, W. L., et al. 2017, \href{https://doi.org/10.3847/1538-4357/834/1/94}{\color{blue}ApJ}\href{https://ui.adsabs.harvard.edu/abs/2017ApJ...834...94S/}{, 823, 94} 

    \bibitem[Tetzlaff et al.(2011)]{Tetal2011} Tetzlaff, N., Neuhäuser, R., \& Hohle, M. M., 2011, \href{https://doi.org/10.1111/j.1365-2966.2010.17434.x}{\color{blue}MNRAS}\href{https://ui.adsabs.harvard.edu/abs/2011MNRAS.410..190T/}{, 410, 190} 

    \bibitem[Tielens(2005)]{T2005} Tielens, A. G. G. M., 2005, The Physics and Chemistry of the Interstellar Medium. Cambridge University Press, New York. 

    \bibitem[Tody(1986)]{T1986} Tody, D., 1986, \href{https://doi.org/10.1117/12.968154}{\color{blue}SPIE}\href{https://ui.adsabs.harvard.edu/abs/1986SPIE..627..733T/}{, 627, 733} 

    \bibitem[Tody(1993)]{T1993} Tody, D., 1993, \href{https://www.aspbooks.org/a/volumes/article_details/?paper_id=9047}{\color{blue}ASPC}\href{https://ui.adsabs.harvard.edu/abs/1993ASPC...52..173T/}{, 52, 173} 

    \bibitem[Vargas-Salazar et al.(2020)]{VSetal2020} Vargas-Salazar, I., Oey, M. S., Barnes, J. R., et al. 2020, \href{https://doi.org/10.3847/1538-4357/abbb95}{\color{blue}ApJ}\href{https://ui.adsabs.harvard.edu/abs/2020ApJ...903...42V/}{, 903, 42} 

    \bibitem[Vink et al.(2015)]{Vetal2015} Vink, J. S., Heger, A., Krumholz, M. R., et al. 2015, \href{https://doi.org/10.1017/S1743921314004657}{\color{blue}HiA}\href{https://ui.adsabs.harvard.edu/abs/2015HiA....16...51V/}{, 16, 51} 

    \bibitem[Walborn(1971)]{W1971} Walborn, N. R., 1971, \href{https://doi.org/10.1086/180754}{\color{blue}ApJ}\href{https://ui.adsabs.harvard.edu/abs/1971ApJ...167L..31W/}{, 167, 31} 

    \bibitem[Walborn(1982)]{W1982} Walborn, N. R., 1982, \href{https://doi.org/10.1086/183747}{\color{blue}APJ}\href{https://ui.adsabs.harvard.edu/abs/1982ApJ...254L..15W/}{, 254, 15} 

    \bibitem[Walborn et al.(2002)]{Wetal2002} Walborn, N. R., Howarth, I. D., Lennon, D. J., et al. 2002 \href{https://doi.org/10.1086/339831}{\color{blue}AJ}\href{https://ui.adsabs.harvard.edu/abs/2002AJ....123.2754W/}{, 123, 2754} 

    \bibitem[Wu et al.(2014)]{Wetal2014} Wu, S.-W., Bik, A., Henning, Th., et al. 2014, \href{https://doi.org/10.1051/0004-6361/201424154}{\color{blue}A\&A}\href{https://ui.adsabs.harvard.edu/#abs/2014A\%26A...568L..13W}{, 568, 13} 

    \bibitem[Zinnecker \& Yorke(2007)]{ZY2007} Zinnecker, H., \& Yorke, H. W., 2007, \href{https://doi.org/10.1146/annurev.astro.44.051905.092549}{\color{blue}ARA\&A}\href{https://ui.adsabs.harvard.edu/abs/2007ARA\%26A..45..481Z/}{, 45, 481} 

\end{thebibliography}
\end{document}